\documentclass[11pt]{article}
\usepackage{amsbsy,amsmath,amsthm,amsfonts,amssymb,amstext,amscd,longtable,flafter,float}

\newtheorem{theo}{\bf Theorem}[section]
\newtheorem{cor}{\bf Corollary}[section]
\newtheorem{defi}{\bf Definition}[section]
\newtheorem{nota}{\bf Notation}[section]
\newtheorem{lem}{\bf Lemma}[section]

\begin{document}

\begin{center}
{\bf {\Large New plans orthogonal through the block factor.}}

\vskip5pt

{\bf {\large Sunanda Bagchi*\\[0pt]
1363, 10th cross, Kengeri Satellite Town, \\
Bangalore 560060, \\
India\\}}
\end{center}

\vskip5pt

\vskip5pt {\bf {\large Abstract  }}
\vskip5pt
In the present paper we construct plans orthogonal through the block factor (POTBs). We describe procedures for adding blocks as well as factors to an initial plan and thus generate a bigger plan. Using these procedures we construct POTBs for symmetrical experiments with factors
having three or more levels. We also construct a series of plans inter-class orthogonal through the block factor for two-level factors.

\vskip10pt

\section{Introduction}
 A situation in which a treatment factor is neither orthogonal nor confounded to a nuisance factor was first explored in Morgan and Uddin (1996) in the context of  nested row-column designs. They derived a sufficient condition for a treatment factor, possibly non-orthogonal to the  nuisance factors, to be orthogonal to another treatment factor. They also derived a sufficient condition for  optimality and constructed several series of orthogonal main effect plans (OMEPs) satisfying optimality properties.  Mukherjee, Dey and Chatterjee (2002) discussed and constructed main effect plans (MEPs) on small-sized blocks, not necessarily orthogonal to all treatment factors. Their plans also satisfy optimality properties. Optimal blocked MEPs of similar type are also constructed in Das and Dey (2004). Wang (2004) constructed plans for two-level factors on blocks of size two, estimating interaction effects also.

 Bose and Bagchi (2007) provided  plans satisfying  properties similar to those of the plans of Mukherjee, Dey and Chatterjee (2002), but requiring fewer blocks. In Bagchi (2010) the concept of orthogonality through the block factor [see Definition \ref {orthbl}] is introduced.
 In that paper it has been shown that a plan orthogonal through the block factor (POTB) may exist in a set up, where an OMEP can not exist.
Making use of the Hadamard matrices in various way,  Jacroux and his co-authors (20011, ... 2017) have come up with a number of such plans, mostly for two-level factors, many of them satisfying optimality properties. Other authors providing POTBs include Chen,  Lin,  Yang,  and Wang (2015) and Saharay and Dutta (2016).

Preece (1966) constructed `BIBDs for two sets of treatments'. Subsequently several authors constructed similar combinatorial
objects. Among these, the ones relevant to the present paper are `balanced Graeco-Latin block designs' of Seberry (1979),
`Graeco-Latin designs of type 1' of Street (1981) and `Perfect Graeco-Latin balanced incomplete block designs (PERGOLAs)' of Rees
and Preece (1999). We  note that all these combinatorial designs are, in fact,  two-factor POTBs  satifying certain additional properties.
We discuss these interesting combinatorial designs briefly in Section 3.

   In the present paper our main objective is to provide plans in those set ups where no OMEP is available, accommodating as many factors as possible and deviating ``as little as possible" from orthogonality.
       We construct a few series of POTBs for symmetrical experiment with factors having three or more levels. We also define
         plans inter-class orthogonal through the block factor (PIOTBs) [see Definition \ref   {PIOTB}] and construct a series of such plans.

      In Section 2 we  present the definition of a POTB along with its  attractive features. The later sections are devoted to construction.
   In Section 3 we obtain a few infinite series of POTBs  for  symmetric experiments with four or less factors, each with five or more levels [see Theorems \ref {POTBk-2,4}, \ref {POTB2} and \ref {POTB2new}]. In Section 4 we describe methods of recursive construction. In Section 5 we use these methods and  construct two series of POTBs  for three-level factors on blocks of size four [see Theorems  \ref  {3-level1} and \ref {3^{(6m+3)N}}]. Finally, in Section 6 we construct an infinite series of PIOTBs with orthogonal classes of small size for two-level factors [see Theorem \ref {PIOTBmn}]. Many of the plans constructed are saturated.

\section{Preliminaries}

We shall consider main effect plans for a symmetrical experiment with $m$  factors,
laid out on blocks of constant size.

\begin{nota} \label{expt}
(a) ${\cal P}$ will denote a main effect plan for a $s^m$ experiment
consisting of $b$ blocks each of size $k$. $n$ will denote the
total number of runs. Thus, $n = bk$.

(b) The set of levels for each factor is   denoted by $S$, the set of integers modulo $s$, unless stated otherwise.
 $S^m$ will denote the following set of $m \times 1$ vectors. $ S^m = \{(x_1, \cdots x_m)' :  x_i \in S \}$.

(c) $A_i$ denotes the $i$th factor, $i = 1,2, \cdots m$.  The vector $x =(x_1,x_2, \cdots x_m)' \in S^m$ represents a
  level combination or run, in which $A_i$ is at level $x_i, i=1,2, \cdots m$.

(d) ${\cal B}  = \{ B_j, \: j=1, \cdots b\}$ will denote the set of
all blocks of ${\cal P}_0$, Thus, $B_j \subset S^m, |B_j| =k, \;1
\leq j \leq b$. Sometimes we describe a plan  in
terms of its blocks.

(e) The replication vector of $A_i$ is denoted by the $s \times 1$
 vector  $r_i$, the $p$th entry of which is the number of runs $x$ of
${\cal P}$ such that $x_i = p, p \in S$. $ R_i$ denotes a diagonal
matrix with  diagonal entries same as those of $ r_i$ in the same order, $1 \leq i \leq
m$.

(f) For $1 \leq i,j \leq m$, the $A_i$ versus $A_j$ incidence matrix
is the $ s_i \times s_j$ matrix $N_{ij}$. The $(p,q)$th entry of
this matrix is  $N^{ij} (p,q)$, which is the number of runs $x$ of
${\cal P}$ such that $x_i = p,\: x_j = q, \; p \in S_i, \; q \in
S_j$. When $j = i, N_{ij} = R_i$.

(g) $L_i$ will denote the $A_i$-versus block incidence matrix, $1
\leq i \leq m$. Thus, the $(p,j)$th entry of the $L_i$ is
$$L^i (p,j) = |{x \in B_j :  x_i = p} |, \; p \in S, 1 \leq j \leq
b,\; 1 \leq i \leq m.$$

(h) The  $ s \times 1$ vector $\alpha^i$ will denote the vector of
unknown effects of $A_i,\: 0 \leq i \leq m$.
\end{nota}

Consider the normal equations for a plan ${\cal P}$ as described above. If we eliminate the general effects and the vector of block effects from this system of equations,  we get the reduced normal equation for the vectors of all (unknown) effects of all the treatment factors. This is a system of $ms$  equations, but it is convenient to view it as  $m$ systems of $s$ equations each, the $i$th system  equations  is of the form
\begin{equation} \label{reducedNE}
\sum_{j=1}^{m} C_{ij;B} \widehat{\alpha^j} = {\mathbf Q}_{i;B}. \end{equation}
Here $C_{ij;B},\; 1 \leq j \leq m$ are the coefficient matrices and  $Q_{i;B}$   is the vector of adjusted (for the blocks) totals for  $A_i$.

For a fixed $i$, we can eliminate $\widehat{\alpha^j}, j \neq i$ from (\ref {reducedNE}) and get
\begin{equation}  \label{reducedNEi} \mbox{ the reduced normal equation for } \widehat{\alpha^i} \mbox{ as } C_{i;\bar{i}} \widehat{\alpha^i} = Q_{i;\bar{i}}.  \end{equation}
We omit the expressions for the quantities $C_{ij;B}, \; C_{i;\bar{i}}, \;Q_{i;B} $ and $Q_{i;\bar{i}}$  above. Those are not necessary here and  are available in Bagchi and Bagchi (2020), for instance. With this background we present a few definitions.

\begin{defi} \label{connected} An m-factor MEP is said to be `connected' if
$Rank(C_{i;\bar{i}}) = s - 1$, for every $i =1,2, \cdots m$.
\end{defi}


\begin{defi} \label{orthbl} [ Bagchi (2010)] Fix $i \neq j \: 1 \leq i,j \leq m$.
The factors $A_i$ and $A_j$ are said to be
 {\bf  orthogonal through the block factor (OTB)} if
\begin{equation}  \label{orthbl-inc}
k  N_{ij} =  L_i (L_j)'.
\end{equation}

We denote this by $A_i \bot_{bl} A_j $.

A plan ${\cal P}$ is said to be a {\bf plan orthogonal through the block
factor (POTB)}  if $A_i \bot_{bl} A_j$ for every pair $(i,j), i \neq
j, i,j =1, \cdots m$.
\end{defi}

{\bf Remark 2.1:} Condition (\ref{orthbl-inc}) is equivalent to equation (7)
of Morgan and Uddin (1996) in the context of nested row-column designs.

Let us try to see the implications of orthogonality through the block factor.  Let  $SS_{i;all}$ (respectively  $SS_{i;B}$) denote  sum of squares for $A_i$, adjusted for all other factors (respectively the block factor). The following results are known.
\begin{theo}\label{orthblCond} Consider a plan ${\cal P}$. Fix $i \in \{1, \cdots m\}$.

(a) [Bagchi(2010)] If for $j \neq i$ $ A_i \bot_{bl} A_j$,  then

(i) $C_{ij;B} = 0$ and (ii)  $Cov(l'\widehat{\alpha^i}, m'\widehat{\alpha^j})=0$, for $l'1_s = 0 = m'1_s$.

(b)[Bagchi (2020)] Further, $ A_i \bot_{bl} A_j,\; \forall j \neq i$ is necessary and sufficient for the following.

(i) $C_{i;\bar{i}} =  C_{ii;B}$ and  (ii) $SS_{i;\bar{i}} = SS_{i;0}$ with probability 1.

 \end{theo}

\vskip5pt
{\bf Discussion :} Theorem \ref {orthblCond} says the following about the inference on the factors of  a connected main effect plan. The inference on a factor  $A_i$ depends only on the relationship between $A_i$ and the block factor  if and only if $A_i$ is orthogonal  to every other treatment factor through the block factor. Moreover, the data analysis of a POTB is very similar to  the data analysis of  a block design with $s$ treatments.

It is well-known that the  orthogonal MEP obtained from an
orthogonal array is the best possible MEP in the sense that the
estimates have the maximum precision among all MEPs in the same
set up. The same cannot be said about an  POTB since its performance
also depends on the relationships of the treatment factors with the
block factor. In the next theorem a guideline for the search for a
`good' POTB is provided. We omit the proof which can be obtained by
going along the same lines as in the proofs of Lemma 1 and Theorem 1
of Mukherjee, Dey and Chatterjee (2002). [See Shah and Sinha (1989)
for definitions, results and other details about standard optimality
criteria]

\begin{theo} \label {optimality} Suppose a connected POTB $\rho^*$
satisfies the following condition. For a factor $A_i$ and a
non-increasing optimality criterion $\phi$, $L_i$ is the
incidence matrix of a block design $d$ which is $\phi$ -optimal in a
certain class of  connected block designs with $s$ treatments and
b blocks of size k each. Then, $\rho^*$ is $\phi$-optimal in a
similar class of  connected m-factor MEPs in the same set-up as
$\rho^*$ for the inference on $A_i$.
\end{theo}

In particular, using the well-known optimality results of Kiefer (1975) and Takeuchi (1961)
 we get the following result.

\begin{cor} \label{u-o OMEP} Suppose $\rho^*$ is a connected POTB. Fix $i \in \{1,\cdots m\}$.

(a) If $L_i$ is the incidence matrix of a BIBD, then, for the inference on
$A_i$, $\rho^*$ is universally optimal in the class of all m-factor connected MEP containing  $\rho^*$.

(b) If $L_i$ is the incidence matrix of a  group divisible design  satisfying $\lambda_2 = \lambda_1 + 1$, then  $\rho^*$ is E-optimal in the class of all m-factor connected MEP containing  $\rho^*$, for the inference on $A_i$.
 \end{cor}

 In view of the above result, we introduce the following term.

\begin{defi} \label{balanced}
A connected POTB is said to be {\bf balanced} if {\bf each of its
factors form a BIBD with the block factor}, that is  $L_i$ is the
incidence matrix of a BIBD for each $i ,\; 1 \leq i \leq m$
\end{defi}

We now present a small example of a balanced POTB on six blocks of size two
each. It has two factors, each with four levels 0,1,2,3.

\vspace{.5em}

{\bf Example 1} [Bagchi and Bose (2007)] :

\vspace{.5em}
\begin{center}
  \begin{tabular}{|ll|cc|cc|cc|cc|cc|cc|}  \hline
              Blocks &$ \rightarrow$&   &  $B_1$     & &$B_2$   &  &  $B_3$  &  &  $B_4$  &  &  $B_5$  &  &  $B_6$  \\
              \hline
Factors $\downarrow$ & $A_1$ & 0 & 2 & 1  & 3 & 0 & 3 & 1 & 2   &0  & 1 &3 &2\\
                     & $A_2$ & 1 & 3 & 0 &  2 & 2 & 1 & 3 & 0   &3  & 2 &0 &1\\\hline \end{tabular}
\end{center}

\section{Construction of plans with a small number of factors}

We shall now proceed to construct POTBs for a symmetric experiment.
 Most of the constructions are of recursive type, in the sense that from a given
initial plan we generate a plan by adding blocks and/or factors.

\begin{defi}\label{AddS} Consider an initial plan ${\cal P}_0$ for an $s^m$ experiment as
described in Notation \ref {expt}. For $B \in {\cal B}$ and $v \in S^m$, $B + v$ will denote the following set of $k$ runs.
$B + v =\{x + v, \; x \in B\}$. Here $x + v = [x_i + v_i : 1 \leq i \leq m]'$,  where the addition in each co-ordinate is modulo $s$.

 By the plan {\bf generated from ${\cal P}_0$ by adding $S$}  we shall mean the plan (for the same experiment)
having the set of blocks $\{ B + u1_m : u \in S, B \in {\cal B} \}$. The new plan ${\cal P}$
will be denoted by ${\cal P}_0 \oplus S$. \end{defi}

We shall now proceed to construction. We begin with plans with a small set of factors.
 Let $ S^+$ denote $S \cup \{\infty\}$. The following rule will define addition in
$S^+$.
 \begin{equation}\label{addInfty} u + \infty = \infty = \infty +u, \; u \in S.\end{equation}

\begin{theo} \label {POTBk-2,4} Suppose $s$ is an integer $ \geq 5$.
Then  POTBs with block size two exists for the following experiments.

(a)   For an $s^2$ experiment a  POTB ${\cal P}$ on $2s$ blocks exists. In the case $s = 5$, ${\cal P}$ is  balanced.

 (b) (i) For an $s^4$ experiment  a  POTB ${\cal P}_1$ on $4s$ blocks exists. If $s = 10$, then
${\cal P}_1$ is E-optimal for the inference on each factor.

(ii) Moreover, if $s \geq 9$, there exists  a  POTB ${\cal P}_2$ with the same parameters  as ${\cal P}_1$, but non-isomorphic to the same. If $s = 9$, ${\cal P}_2$ is  balanced.

 (c) A POTB ${\cal P}$ for a $(s+1)^4$  experiment with $6s$ blocks exists, whenever $ n \geq 7$.
 \end{theo}

{\bf Proof :}  In each case, we present the blocks of an initial plan ${\cal P}_0$.
The required plan is ${\cal P}_0 \oplus S$  [see Definition \ref {AddS}]. Here $a,b,c,d$ are distinct members of  $S \setminus \{0\}$.
That the final plan is a POTB  can be verified by straightforward computation. Proofs for the optimality properties are presented.

(a) The blocks of ${\cal P}_0$ are given below.

 \vspace{.5em}
 \begin{center}
\begin{tabular}{|ll|cc|cc|}  \hline
              Blocks &$ \rightarrow$&   &  $B_1$     & &$B_2$   \\
              \hline
Factors $\downarrow$ &$A_1$ &      a         & -a   &    b         &  -b    \\
                     &$A_2$ &      b         &  -b  &    -a         & a
\\\hline\end{tabular}.
\end{center}
\vspace{.5em}

If $s=5$, taking $a=1,b=2$ we get a balanced POTB.

\vspace{.5em}

 (b) (i) The blocks $B_l, \; l = 1, \cdots 4$ of ${\cal P}_0$  are as follows.

 \vspace{.5em}
 \begin{center}
 \begin{tabular}{|ll|cc|cc|cc|cc|}  \hline
              Blocks &$ \rightarrow$&   &  $B_1$     & &$B_2$   &  &  $B_3$  &  &  $B_4$    \\
              \hline
Factors $\downarrow$ &          $A_1$  &  0  & a &a    &-a & 0   &b& -b & b \\
       &   $A_2$  & a   &-a &0   &-a &  -b & b  & 0 & b \\
        &  $A_3$ &  0   & b & b   &-b &  -a & 0 &  a & -a\\
        &  $A_4$ &  b   &-b &  0   &-b &  a   &-a  & a & 0 \\ \hline
\end{tabular}.
\end{center}

\vspace{.5em}

 If $n=10$, we take $a=1$ and $b=3$. Then for every $i =1, \cdots 4$,
 $L_i$ is the incidence matrix of a group divisible design with five
 groups, the jth group being the pair of levels $\{j, j+5\} \:
 j=0,\cdots 4$, satisfying $\lambda_1 = 0$ and $\lambda_2 = 1$. This plan
  is, therefore, E-optimal for the inference on all the four factors by the result of
  Corollary \ref {u-o OMEP} (b).

\vspace{.5em}

(b) (ii) The  blocks  $B_l, \; l = 1, \cdots 4$  of ${\cal P}_0$
 are as follows.
\vspace{.5em}
 \begin{center}
\begin{tabular}{|ll|cc|cc|cc|cc|}  \hline
              Blocks &$ \rightarrow$&   &  $B_1$     & &$B_2$   &  &  $B_3$  &  &  $B_4$   \\
              \hline
Factors $\downarrow$ &       $ A_1$   & a  & -a &  b  & -b &  c & -c &  -d & d \\
      & $ A_2 $  & b  & -b &  -a & a  & -d & d & -c & c \\
      & $ A_3 $  & c & -c  &  d  & -d  &-a & a &  b  & -b \\
      & $ A_4 $    & d & -d  &  -c & c &   b  & -b &a  & -a  \\\hline
        \end{tabular}
\vspace{.5em}
\end{center}
By taking $a=1,b=2,c=3$ and $d=4$ in the case $s = 9$, we get a
balanced POTB.

\vspace{.5em}

(c) The set of levels for each factor is $S^+$. The  blocks  $B_l, \; l = 1, \cdots 6$ of the initial plan are as follows.

\vspace{.5em}

 \begin{center}
\begin{tabular}{|ll|cc|cc|cc|cc|cc|cc|}  \hline
              Blocks &$ \rightarrow$&   &  $B_1$     & &$B_2$   &  &  $B_3$  &  &  $B_4$  &  &  $B_5$  &  &  $B_6$  \\
              \hline
              Factors $\downarrow$ &
  $A_1$   &0 & $\infty$ &a  & -a  & b & -b & c & -c & a  & -a & a &-a\\
  & $A_2$   &a  & -a &0 &$\infty$ & c & -c  & -b & b  &a  & -a &-a &a\\
  & $A_3 $  & b & -b & c & -c   &0 &$\infty$ &a  & -a &-c & c &-c & c\\
  & $A_4$   &c & -c  &b & -b    &a & -a    &0&$\infty$ &-c & c  &c &-c\\
\hline \end{tabular} . $\Box$
\end{center}
\vspace{.5em}

 We now list a few combinatorial structures in the literature which are actually balanced POTBs (for symmetrical or asymmetrical experiments).

(a) {\bf Balanced  Graco-Latin block design} defined and constructed
in Seberry (1979) heve two factors.

(b) {\bf  Graco-Latin block design of type 1} of  Street (1981) are
also two-factor  balanced POTBs satisfying $ {\mathbf N}_{12} = J.$

 (c)  {\bf Perfect Graeco-Latin balanced incomplete block designs
 (PERGOLAs)} defined and discussed extensively in Rees and Preece (1999)
are  two-factor  balanced POTBs satisfying

\begin{equation} \label{Pergola}
{\mathbf N}_{12} {\mathbf N}'_{12} =
{\mathbf N}'_{12} {\mathbf N}_{12} = f I_s + g J_s, \; \mbox{ where f, g
are integers}.\end{equation}
 Here $I_n$ is the identity matrix and $J_n$ is the
 all-one matrix of order $n$.

 (d) {\bf Mutually orthogonal BIBDs} defined and constructed by Morgan and Uddin
 (1996) are multi-factor  balanced POTBs.

\vspace{.5em}

{\bf Remark 3.1:} The definition of neither  balanced  Graco-Latin
block designs nor of  mutually orthogonal BIBDs include condition
(\ref {Pergola}). However, it is interesting to note that all
these designs constructed so far do satisfy this condition.  One would,
therefore, suspect that this condition is implicit in the
definition. We have, however, found a balanced POTB which does not
satisfy this condition, as is shown below.

\begin{theo} \label{POTB2} Let $s$ be a positive integer $\geq 5$.
 Then

 (a) there exists a symmetric POTB  ${\cal P}$ with three factors each
having $s+1$ levels on $ b = 6s$ blocks of size two.

(b) In the case $s = 5$, we get a Balanced POTB. The restriction to any two of the factors reduces it to a PERGOLA, except that
 condition (\ref {Pergola}) is not satisfied.
\end{theo}

{\bf Proof :} (a) Let $S^+$ be the set of levels for each factor. Consider
an initial plan ${\cal P}_0$ with the set of factors $\{A_0,A_1, A_2\}$ and  ${\cal B} = \{B_{ij},\; i=1,2,\:j = 0,1,2\}$, where $B_{ij}$'s are  as shown in the table below.
The required plan ${\cal P} = {\cal P}_0 \oplus S$.

     \begin{center}
\begin{tabular}{|ll|cc|cc|cc|cc|cc|cc|}  \hline
              Blocks &$ \rightarrow$& &$ B_{10}$ & &$ B_{11}$ & &$ B_{11}$ & &$ B_{20}$   & &$ B_{21}$   & &$ B_{22}$       \\
              \hline
              Factors $\downarrow$ &
$A_0$   & $\infty$   & 0  & -1 & 1 & 0 & 1 & $\infty$ & 0 & 1 & 2 & 0 & 2 \\
&$A_1 $  &   0  & 1  &$\infty$ & 0 & -1 & 1 & 0 & 2  & $\infty$ & 0 &  1  & 2  \\
&$A_2$   &  -1  & 1  & 0 & 1 &$\infty$ & 0 &  1 & 2 & 0 & 2& $\infty$ & 0 \\\hline
\end{tabular}.
\end{center}

 That ${\cal P}$ satisfies (\ref {orthbl-inc}) follows by
 straightforward verification.

(b) Let $s = 5$.  One can verify that the incidence matrices
satisfy the following.

 \begin{equation} \label{5incMat}
 N_{ij} =  \left [ \begin{array} {cccccc}
0 & 2 & 2 & 2 & 2 & 2  \\
2 & 2 & 2 & 1 & 1 & 2 \\
2 & 2 & 2 & 2 & 1 & 1\\
2 & 1 & 2 & 2 & 2 & 1 \\
2 & 1 & 1 &  2 & 2 & 2 \\
2 & 2 & 1 & 1& 2 & 2 \\\end{array}  \right ],\; i,j = 0,1,2 . \end{equation}
 \begin{equation} \label{5incMatBl}
\mbox{Moreover, }  L_{i} (L_{i})' = 8I_6 + 2 J_6,\: i=0,1,2.\end{equation}

We see that each $L_i$ is the incidence matrix of a BIBD with parameters $(v = 6, b = 30, r = 10, k = 2, \lambda = 2)$. Thus, by Definition \ref {balanced}  ${\cal P}$ is a balanced POTB. However,
$N_{ij}$ does not satisfy (\ref {Pergola}), $ i \neq j, \; i,j = 0,1,2$.
 $\Box$

\vspace{.5em}

Next we construct a series of balanced POTBs using finite fields.
 We first  introduce the following notation.

\begin{nota} \label{Unionnota}

(i) $\bigsqcup$ denotes an union counting multiplicity.

(ii) For a set $A$ and an integer $n$,  $nA$ denotes the multiset in which every member of $A$ occurs $n$ times.

(iii) For subsets A and B of a group $(G,+)$,
$$ A - B = \{ a-b : a \in A, b \in B \}.$$
\end{nota}

\begin{nota} \label{GFnota}(i) $s$ is  an odd prime power. $t = (s-1)/2$. $F$ denotes the Galois field of
order $s$. Further, $F^* = F \setminus \{0 \}$ and $F^+ = F \cup \{\infty\}$.

(ii) $\alpha$  denotes a primitive element of $F$.

 (iii) $C_0$ denotes the subgroup  of order t of the multiplicative group of $F$ and $C_1$ the coset of $C_0$. Thus, $C_0$ is
 the set of all non-zero squares of $F$, while $C_1$ is the set of all non-zero non-squares of $F$.



(iv) $(i,j) $ denotes the number of ordered pairs of integers (k,l) such
that the following equation is satisfied in $F$. [ This notation is
borrowed from the theory of cyclotomy]

 $$ 1 + \alpha^k = \alpha^l ,\; k \equiv i, l \equiv j \pmod 2.$$
 \end{nota}

 We present the following well-known result for ready reference. [See equations (11.6.30), (11.6.40) and (11.6.43) of Hall
(1986)].

\begin{lem} \label{cyclotomy}
 The  difference between the cosets  of $F^*$ can be expressed in terms of the cyclotomy numbers  as follows.
$$ C_1 - C_0 = \bigcup\limits_{k=0}^{1} (k,1) C_k.$$
The following cyclotomy numbers are known.

{\bf Case 1: t odd.} (0,0) = (1,1) = (1,0) = (t-1)/2, (0,1) =
(t+1)/2.

{\bf Case 2: t even.} (0,0) = t/2 -1, (0,1) = (1,0) = (1,1) = t/2.
\end{lem}

\vspace{.5em}

{\bf A series of two-factor balanced POTBs :}

\begin{theo} \label{POTB2new} Suppose $s$ is an odd prime or a prime
power. Then there exists a balanced POTB ${\cal P}^*$ for a $(s+1)^2$ experiment
on $b = 2s$ blocks of size $(s+1)/2$ . \end{theo}

{\bf Proof :}  The set of levels of each factor is $F^+$. We shall present the initial plan ${\cal P}_0$ consisting of a pair of blocks.
The required POTB is  ${\cal P}^* = {\cal P}_0 \oplus F$.

Let $\delta \in C_1$. Consider three $2 \times (t+1)$ arrays $R^0, R^1$ and $R^2$, the rows of which are indexed by $\{0,1\}$ and the columns by
$C_0 \cup \{0\}$. The entries of the arrays are as given below.
\begin{eqnarray} \label{blockarray}
R^0(1,0) = R^1(0,0) = R^2 (0,0) = 0 \mbox{ and }  R^0(0,0) = R^1(1,0) = R^2 (1,0) = \infty. \\
\mbox{For } x = 0,1, \; y \in C_0, \;  R^0(x,y) = \delta^x y,  R^1(x,y) = \delta^{-x} y \mbox{ and }  R^2(x,y) = \delta^{x-1} y. \end{eqnarray}
For $i = 0,1,2$, let $B_i$ be the block, the runs of which are the columns of $R^i$. When $t$ is even,
$B_0$ and $B_1$ constitute ${\cal P}_0$,   while $B_0$ and $B_2$  constitute ${\cal P}_0$  when $t$ is odd.

Clearly block size is $t+1 = (s+1)/2$. To show that ${\cal P}^*$ satisfies the required property, we have to show that

(a) ${\cal P}^*$ is a POTB and (b) each factor  forms a BIBD with the block factor.

Condition (b) follows from the construction in view of Lemma \ref {cyclotomy}. So, we prove (a). Let us write $N$ for $N_{12}$. The rows and
columns of $N$ are indexed by $F^+$. From (\ref {blockarray}) and (3.9), we see that
\begin{equation} \label{incMatPart} N (ii) = 0, \;i \in F^+ \mbox{  and } N (\infty,i) =  N (i,\infty) =1, \;i \in F. \end{equation}

So, we assume $i \neq j \in F$. Let $u = j-i$. Then, $N (ij)$ is the number of times $u$ appears in the multiset
$$ \left \{\begin{array} {ll}
(\delta -1) C_0 \bigsqcup (\delta^{-1} -1) C_0 & \mbox{ if t is
even}\\
(\delta -1) C_0 \bigsqcup (1 - \delta^{-1}) C_0 & \mbox{ if t is
 odd} \end{array} \right . $$
Since $-1 \in C_0$ if and only if $t$ is even, $\delta^{-1} -1$ is in the same coset as $\delta -1$ if and only if $t$ is odd. Therefore, the relations above together with (\ref {incMatPart})  above imply that
\begin{equation} \label{IncMatrN} N = J_{s+1} - I_{s+1}. \end{equation}

Now we take up $L_1 L'_2 = H$ (say). From (\ref {blockarray}) and (3.9), we see that
\begin{equation} \label{L1L2Part} H (ii) = 0, \;i \in F^+ .\end{equation}
Further, for every $i \in F$, $H (\infty,i)$  is the replication number of $i$ in the block design generated by the initial block $\{0\} \cup
C_1$. Similarly, $H (i,\infty)$ is the replication number of $i$ in the block design generated by the initial block $\{0\} \cup C_0$ if
$t$ is odd and $\{0\} \cup C_1$ otherwise. Thus,
\begin{equation} \label{L1L2Part2}
H (\infty,i)  = H (i, \infty) = t+1, \; i \in F. \end{equation}

We, therefore, assume $i \neq j,\; i,j \in F$. Let $u = j-i$. Then,
$H (ij)$ is the number of times $u$ appears in the multiset
$$ \left \{\begin{array} {ll}
((\{0\} \cup C_1) - C_0) \bigsqcup (C_1 - (\{0\} \cup C_0)) & \mbox{ if $t$ is
even}\\
((\{0\} \cup C_1) - C_0) \bigsqcup (C_0 - (\{0\} \cup C_1))  & \mbox{ if $t$ is
odd} \end{array} \right . $$
These relations, together with (\ref {L1L2Part}), (\ref {L1L2Part2}) and  Lemma \ref {cyclotomy} imply that $H = (t + 1) (J_{s+1} - I_{s+1})$. Therefore, in view of (\ref {IncMatrN}), (\ref {orthbl-inc}) follows and we are done. $\Box$

\section{More on recursive construction}

In this section we describe  procedures for adding factors as well as blocks to an initial plan.

\begin{nota}\label{AddBl} Consider a subset $V$ of $S^m$.

  For every $i, 1 \leq i \leq m$, $V_i$ will denote the following
multiset of $|V|$ members of $S$. $V_i  = \{v_i :  v = (v_1, \cdots v_m)'
\in V \}$. Similarly, $V_{ij}$ will denote the following
multiset of $|V|$ members of $S \times S$. $V_{ij}  = \{ (v_i,v_j) :  v = (v_1, \cdots v_m)'
\in V \}$.

\end{nota}

\begin{defi} \label{AddingBlock}
Consider an initial plan ${\cal P}_0$ for an $s^m$ experiment as
described in Notation \ref {expt}.  Let $V$ be as in Notation
\ref {AddBl}. By the plan ${\cal P}_0 + V$ {\bf generated from ${\cal
P}_0$ along $V$} we shall mean the plan (for the same experiment)
having the set of blocks ${\cal B} + V  = \{B + v : v \in V, B \in
{\cal B} \}$, where $B + v$ is as in Definition \ref  {AddS}. Usually, $V$ will contain the $0$-vector, so that the
blocks of ${\cal P}_0$ will also be blocks of ${\cal P}_0 + V$.
\end{defi}

\vspace{.5em}

The next lemma provides a few sufficient conditions on ${\cal P}_0$ and $V$ so
that a given pair of factors are orthogonal through the block factor  in  ${\cal P}_0 + V $. The proof is by direct verification.

\vspace{.5em}

{\bf Remark 4.1:} In an initial plan, say ${\cal P}_0$, one or more
levels of one or more factors may be absent. ${\cal P}_0$ may still be
a POTB if (\ref {orthbl-inc}) holds (with one or more row/column of
 $N_{ij}$'s  being null vectors) for every unordered pair of
$(i,j)$. In such cases one has to choose $V$ such that all levels of
all factors do appear in ${\cal P}_0 + V$.

\begin{lem}\label{suffPOTB} Consider an initial plan ${\cal P}_0$
 for an $s^2$ experiment. For $V \subset F \times F$, consider   ${\cal P}_0 + V $.  The following
conditions on ${\cal P}_0$ and $V$ are sufficient for ${\cal P}_0 + V$ to
be a POTB.

(a) In  ${\cal P}_0$ all the levels of the first factor appear and $V
= \{(0,i),\: i \in S \}$.

(b) ${\cal P}_0$ is arbitrary and $V = \{(i,j),\: i,j \in S \}$.

(c) ${\cal P}_0$ has a pair of  blocks $B_0,B_1$ each of size 2, as
described below. Let $i \neq j,\; k \neq l \in S$. Let $x_0 = (i,i)',\: y_0 =
(j,j), \: x_1 = (k,l)'$ and $y_1 = (l,k)'$. $B_i$ consists of runs
$x_i$ and $y_i$, $i = 0,1$. $V = (u,u),\: u \in S$.

(d)  ${\cal P}_0$ is a POTB in which with one or more levels of one or
both factors may be absent. $V$ is such that every member of $S$
appears at least once in each $V_i, i = 1,2$.
\end{lem}


Our next procedure enlarges the set of factors of a given plan, while keeping the
number of blocks fixed.

\begin{defi}\label{powerPlan} (a) Consider a plan ${\cal P}$
as in Notation \ref {expt}. Suppose there is another
plan  $ {\cal P}'$ having $b$ blocks of size $k$ each.
We shall combine these two plans to get
 another one with a larger set of factors.

Let $x_{ij}$ (respectively $\tilde{x}_{ij}$)
denote the $j$th run in the ith block of ${\cal P}$ (respectively  $
{\cal P}'$) ,$\: 1 \leq j \leq k, \: 1 \leq i \leq b$. Let $y_{ij} =
\left [\begin{array}{cc} x_{ij} & \tilde{x}_{ij}\end{array} \right ]', \:1
\leq j \leq k, \: 1 \leq i \leq b$. Then, the plan on b blocks of
size $k$ with $y_{ij}$ as the $j$th run in the ith block, $1 \leq j
\leq k, \: 1 \leq i \leq b$ is said to obtained by joining the
factors of ${\cal P}$ and  $ {\cal P}'$ together. The new plan
will be denoted by $\left [\begin{array}{cc}{\cal P} & {\cal P}'
\end{array} \right ]$.

(b) In case $ {\cal P}'$ is a copy of  ${\cal P}$ then $\left
[\begin{array}{cc}{\cal P} & {\cal P}' \end{array} \right ]$ is
denoted by ${\cal P}^2$. For $t \geq 3$, the plan ${\cal P}^t$ is
defined in the same way. In this case we name the factors of ${\cal
P}$ and its power ${\cal P}^t$ as in the notation below.
\end{defi}

\begin{nota}\label{allFactors} Consider a plan  ${\cal P}$ having
a set of $m$ factors ${\cal F}_0 = \{A, \cdots M\}$. The set of
factors of ${\cal P}^t$ will be named as
$${\cal F} = \bigcup\limits_{i = 1}^t {\cal F}_i, \mbox{where }
 {\cal F}_i = \{A_i, \cdots M_i \}.$$

\end{nota}


Combining Definitions \ref {AddingBlock} and \ref {powerPlan} we get
 a recursive construction described below.

\begin{defi} \label{recursiveConstr}
 Consider an initial plan ${\cal P}_0$ for an $s^m$ experiment laid on $b$ blocks
 of size $k$ each. Consider a $p \times q$ array $H = ((h_{ij}))_{1
\leq i \leq p, 1 \leq j \leq q}$. We now obtain a plan for an
$s^{mq}$ experiment on $bp$ blocks of size $k$ using the array $H$
as follows. We first  obtain ${\cal P}_0^q$ following Definition
\ref {powerPlan}.

Let $v_i = \left [ \begin{array}{ccccccc} h_{i1}.1^{\prime }_t &
h_{i2}.1^{\prime }_t & \cdots & h_{iq}.1^{\prime }_t \end{array}
\right ]^{\prime }, \: 1 \leq i \leq p$ and $V_H = \{v_i, \: 1 \leq
i \leq p\}$.

Our required plan ${\cal P}$  is ${\cal P}^q_0 + V_H$ and it
will be denoted by $H \Diamond {\cal P}$. Symbolically,
\begin{equation}  \label{NewPlan}
{\cal P} = H \Diamond {\cal P}_0 = {\cal P}_0^q + V_H.
\end{equation} \end{defi}

 Our task is to find a suitable array $H$ so that the plan $H \Diamond {\cal P}_0$ satisfies certain desirable properties. A natural choice
for $H$ is an orthogonal array of strength $2$. We shall use a modification of an orthogonal array so as to accommodate a few more factors.

\begin{defi} [Rao(1946)]
\label{OAnmst} Let $m,N,t \geq 2$ be integers and $s$ is an integer
$\geq 2$. Then an orthogonal array of strength $t$ is an $m \times
N$ array, with the entries  from a set $S$ of $s$ symbols satisfying
the following. All the $s^t$ $t$-tuples with symbols from $S$ appear
equally often as columns in every $t\times N$ subarray.  Such an
array is denoted by $OA(N,m, s, t)$.
\end{defi}

\begin{nota}\label{ExtOA} (a) The set of symbols of an $OA(N,m,s,2)$
is assumed to be the set of integers modulo $s$.

(b) The array obtained by adding a column of all zeros (in the $0$th
position, say) to an $OA(N,m-1,s,2)$ will be denoted by $Q(N,m,s)$.
\end{nota}

\vspace{.5em}

Exploring the properties of an orthogonal array of strength $2$, we
get the following result from the recursive construction described
in Definition \ref {recursiveConstr}.

\begin{theo} \label{PlanThruOA} Consider a plan ${\cal P}_0$ for an
$s^t$ experiment on $b$ blocks of size $k$ each. If an
$OA(N,m-1,s,2)$ exists, then $\exists$ a plan ${\cal P}$ with a set
of $s^{mt}$ factors on $bN$ blocks of size $k$ each with the
following properties. Here the  factors of ${\cal P}_0^q$ as well as ${\cal P}$ are named according to   Notation \ref {allFactors}.

(a) For $P \neq Q, P,Q \in {\cal F}_0$,  $P_i \bot_{bl} Q_i $ for every
$i,\: 0 \leq i \leq m - 1$, if and only if $P \bot_{bl} Q $ in ${\cal
P}_0$.

(b) $P_i \bot_{bl} Q_j , \; P,Q \in {\cal F}_0,\; i \neq j, 0 \leq i,j \leq
m -1 , $. \end{theo}

{\bf Proof :} By assumption $Q = Q(N,m,s)$ exists. The required plan  ${\cal %
P}$ is $Q \Diamond {\cal P}_0$. Property  (a) follows from the
construction while (b) follows from (b) of Lemma \ref{suffPOTB}.

\vspace{.5em}

{\bf Remark  4.2 :} Table 1 of Rees and Preece (1999) presents a number of examples of PERGOLAs [see the statement proceeding (\ref {Pergola})]. An Application of Theorem \ref {PlanThruOA} on each of them would yield a balanced POTB for a larger set of factors.

\vspace{.5em}

    Finally, we describe a procedure of modifying the sets of levels of factors.
Specifically, given a pair of plans with the same number of factors and the
same block size, we obtain a plan by merging the sets of levels of
the corresponding factors of the  given plans.

\begin{defi}\label{combPlans} Consider a pair of plans ${\cal P}_1$ and ${\cal
P}_2$ each having $t$ factors and blocks of size $k$. Let $S_i$
denote the set of levels of each factor of ${\cal P}_i, \; s_i =
|S_i|, \; i = 1,2$. We assume that $S_1 \neq S_2$.
Let $U = S_1 \cup S_2$ and $u = |U|$. The plan consisting  of all the blocks of
${\cal P}_1$ and  $ {\cal P}_2$ taken together will be viewed as a plan, say
${\cal P}_1 \cup {\cal P}_2$, for an $u^t$  experiment in the following sense.

(a) Each factor of  ${\cal P}_1 \cup {\cal P}_2$ will have $U$ as the set of levels.

(b) Fix $p \in U$. Let ${\cal R}^{ij}_p$ denote the set of runs of ${\cal P}_j$, in which the
  level $p$ of the $i$th factor  appears, $j = 1,2, 1 \leq i \leq t$.
   [Needless to mention that ${\cal R}^{ij}_p = \phi$ if $p$ is not in $S_j$.]
 Then, the level $p$ of the  $i$th  factor of ${\cal P}_1 \cup {\cal P}_2$ appears in exactly the runs in
   ${\cal R}^{i1}_p \sqcup {\cal R}^{i2}_p, 1 \leq i \leq t$.
   \end{defi}

 {\bf Remark 4.3:} From Definition \ref  {combPlans} we see that
  for $p \in U$, the replication number of level $p$ of the $i$th
factor of  ${\cal P}_1 \cup {\cal P}_2$ is $r^{i1} (p) + r^{i2} (p)$, where
$r^{ij} (p)$ is the replication number of level $p$ of the $i$th
factor of  ${\cal P}_j$.

  For instance, in Theorem \ref {3-level1} below, Definition \ref  {combPlans}
  is used to construct ${\cal P}_h$  by merging the corresponding factors of  ${\cal P}_{1h}$
 and ${\cal P}_{2h}$. There, $S_1 = \{0,1\}$, while $S_2 = \{0,2\}$. Thus, while both  ${\cal P}_{1h}$
 and ${\cal P}_{2h}$ are equireplicate, the replication number of level $0$ of each factor of ${\cal P}_h$
 is double of the levels $1$ and $2$ of the same factor.

The following result is an immediate consequence of Definition \ref{combPlans} .

\begin{lem} \label{unionPlan} Consider a pair of connected plans ${\cal P}_1$ and ${\cal
P}_2$, as in Definition \ref {combPlans} (recall Definition \ref {connected}).
Then, we can say the following about the plan ${\cal P}=  {\cal P}_1 \cup {\cal P}_2$.

(a) If both ${\cal P}_1$ and  $ {\cal P}_2$ are POTB, then so is
${\cal P}$.

(b) ${\cal P}$ is connected,
if and only if $S_1 \cap S_2 \neq \phi$. \end{lem}

\section{Construction of  POTBs for three-level factors}

In this section we make use of the tools described in Section 4 to generate
plans for  three-level factors. The factors of the initial and
final plans are named in accordance with  Notation \ref {allFactors}.

\begin{theo}\label{3-level1} If $h$ is the order of a Hadamard matrix,
  then there exists a connected and saturated POTB ${\cal P}_h$ for a $3^{3h}$
experiment in $2h$ blocks of size $4$ each.
 \end{theo}

{\bf Proof :} Let $O_4 = OA(4,3,2,2)$ with $S= \{0,1\}$. Let $ {\cal P}_0$ be the plan consisting of a single block
consisting of the four columns of $O_4$ as runs. Thus, $ {\cal P}_0$ is an OMEP for a $2^3$ experiment.

By hypothesis $Q = Q(h,h,2)$ exists. Let ${\cal P}_{1h} = Q \Diamond {\cal P}_0$ and ${\cal P}_{2h}$ be obtained from ${\cal P}_{1h}$ by replacing level 1 of every factor by the level 2. Next we construct our required plan
${\cal P}_h = {\cal P}_{1h} \cup {\cal P}_{2h}$ by using Definition \ref {combPlans}.
 By construction ${\cal P}_h$  has $2h$ blocks of size $4$ each.

We now show that ${\cal P}_h$ is a POTB. We note that by Theorem
\ref{PlanThruOA},  each of ${\cal P}_{1h}$ and ${\cal P}_{2h}$ is a
POTB for a $2^{3h}$ experiment on $h$ blocks of size 4 each. The
sets of levels of each factor of them are $\{0,1\}$ and $\{0,2\}$
respectively. It follows from  Lemma \ref {unionPlan} that ${\cal
P}_h$ is a connected POTB for an experiment with $3h$ factors, the
set of levels of each factor being $\{0,1,2\}$. Since the available
degrees of freedom for the treatment factors is $2h(4-1)$ which is
the same as the required degrees of freedom, the plan is saturated.
$\Box$

We now take $h = 2$ and present the plan ${\cal P}_2$ for a $3^6$
experiment on four blocks of size four each.

\vspace{.5em}
 \begin{center}
{\bf  Table 5.1 : The plan $ {\cal P}_2$}

\vskip5pt

\begin{tabular}{ll|c|c|c|c}
Blocks &$ \rightarrow$ &   $B_{01}$  &$B_{02}$ & $B_{11}$ &  $B_{12}$ \\
 \hline
 Factors $\downarrow$    &$A_1$& 00  11 & 00  11 & 00  22 & 00  22 \\
 &$B_1$& 01  01 & 01  01 & 02  02 & 02  02\\
             &$C_1$& 01  10 & 01  10 & 02 20  & 02  20  \\
             &$A_2$& 00  11 & 11  00 & 00 22 & 22  00 \\
             &$B_2$& 01  01 & 10  10 & 02  02 & 20 20\\
             &$C_2$& 01  10 & 10  01 & 02  20 & 20  02 \\\hline
            \end{tabular}
 \end{center}

\vspace{.5em}

For the next construction we need some more notations.

\begin{nota} $O_4$ is as in the proof of Theorem \ref {3-level1}.
$T_4$ will denote the array obtained from $O_4$ by replacing each 1
by 2 and $\tilde{T_4}$  the array obtained from $T_4$ by
interchanging 0 and 2. \end{nota}

\begin{theo}\label{3^3} A POTB  for a  $3^3$
experiment on two blocks of size four exists. \end{theo}

{\bf Proof : } Let  $B_{10} = O_4$,  $B_{20} = T_4$  and  $B_{02} = \tilde{T}_4$. The set of columns of each of them constitutes an OMEP for a $2^3$ experiment, the set of levels of factors being $\{0,1\}$ for $B_{10}$, while $\{0,2\}$ for the other two.

Let $\rho_{1}$ (respectively $\rho_2$) denote the plan consisting of the pair of blocks $B_{10}, B_{20}$ (respectively
$B_{10}, B_{02}$). By Lemma \ref {unionPlan}, each of $\rho_1$ and $\rho_2$ is a POTB for a $3^3$ experiment.$\Box$

Using the pair of plans constructed above, we generate a bigger plan.

\begin{theo}\label{3^{(6m+3)N}} (a) If there exists an $OA(N,m,3,2)$, then
 there exists a connected POTB ${\cal P}_m$ for a $3^{3(2m+1)}$
experiment in $2N$ blocks of size $4$ each.

 In particular ${\cal P}_m$ is  saturated whenever $N = 3^n$ and
$m = (3^{n-1} -1)/2$, for an integer $n \geq 2$.

(b) There exists a connected POTB for a $3^9$ experiment in $6$ blocks of size $4$ each.
 \end{theo}

{\bf Proof of (a):} Let the factors of $\rho_1$ and $\rho_2$ be named as
$A,B,C$ and $\tilde{A}, \tilde{B}, \tilde{C}$ respectively. Let $O =
OA(N,m,3,2)$ and $Q = Q(N,m,3)$. We now  use Definition \ref {combPlans}
to generate bigger plans
${\cal P} _{1m}$ and ${\cal P}_{2m}$ as follows.
\[{\cal P}_{1m} = Q \Diamond \rho_{1} \mbox { and }  {\cal P}_{2m} = O
\Diamond \rho_2 .\]

 Clearly, ${\cal P}_{1m}$ and ${\cal
P}_{2m}$ are plans for $3^{3(m+1)}$ and  $3^{3m}$ experiments
respectively, each on $2N$ blocks of size 4. Following Notation
\ref {allFactors}, we name of the factors of these plans as follows.
\begin{eqnarray*} \mbox{The factors of } {\cal P}_{1m}  \mbox{ are } &
A_0,B_0,C_0, A_1,B_1,C_1, \cdots A_{m}, B_{m}, C_{m} \\
\mbox{ and the factors of } {\cal P}_{2m}  \mbox{ are } &
\tilde{A}_1,\tilde{B}_1, \tilde{C}_1, \cdots \tilde{A}_m,
\tilde{B}_m, \tilde{C_m}.
\end{eqnarray*}

 Now we  combine the factors of  ${\cal P}_{1m}$ and ${\cal P}_{2m}$ following  Definition \ref {powerPlan} (a) and thus obtain our
 required plan ${\cal P}_m$. Symbolically,
$${\cal P}_m = \left [ \begin{array}{cc}
{\cal P}_{1m} & {\cal P}_{2m}%
\end{array}
\right ].$$

 By construction, ${\cal P}_m$ is a plan for $2m+1$ three-level factors on $2N$ blocks of size 4 each. We shall
now show that it is a POTB.

 Theorems \ref {PlanThruOA} and \ref {3^3} imply that each one of ${\cal P}_{1m}$ and ${\cal P}_{2m}$ is a POTB. Therefore, if we  show the following relation, then we are done.
\begin{equation}\label{PQorthogonal}
P_i \bot_{bl}  \tilde{Q}_j ,\;  P,Q \in \{A,B,C\},\; i \in I \cup \{0\}, \; j\in I, \mbox{ where }I = \{1, \cdots m\}. \end{equation}

To show this relation, we fix $P_i$ and $\tilde{Q}_j$ as above.

{\bf Case 1. $i ,j \in I$ :}  Since $\rho_1$ and $\rho_2$ are POTBs,  (\ref {PQorthogonal})  follows from  Lemma \ref {suffPOTB} (d), whenever  $Q \neq P$. Again, (c) of the same Lemma proves (\ref {PQorthogonal}) for the case $Q = P$.

{\bf Case 2. $i = 0, \; j \in I$ :}  We take $P_0$ as the first and $\tilde{Q}_j$ as the second factor. Then applying
 Lemma \ref {suffPOTB} (a) we get (\ref {PQorthogonal}).

Hence the proof of the first part is complete.

\vspace{.5em}

To prove the second part, we see that ${\cal P}_m$ is saturated when $N = 2m+1$.
Now Rao (1946) has shown that an $OA(s^n, (s^n - 1)/(s-1), s,2)$
exists whenever $n \geq 2$. (see Theorem 3.20 of Hedayat, Sloane and
Stufken (1999) for instance). Putting $s = 3$, we get  the result.

{\bf Proof of (b) :} Let $O =  \left [ \begin{array}{ccc} 0 & 1 & 2 \\
                                                          0 & 2 & 1 \\\end{array} \right ].$
and $Q = \left [ \begin{array}{ccc} 0 & 0 & 0 \\ 0 & 1 & 2 \\
                                                          0 & 2 & 1 \\\end{array} \right ].$
 Now the construction for the plan, say ${\cal P}_1$,  is just like that in Case (a). The verification is also exactly like the same in Case (a) with $I = \{1\}$.
$\Box$

\vspace{.5em}

 We now present ${\cal P}_1$.

 \begin{center}
{\bf  Table 5.2 : The plan $ {\cal P}_1$}

\vskip5pt

\begin{tabular}{ll|c|c|c|c|c|c}
Blocks &$ \rightarrow$ & $B_{10}$  & $B_{20}$ & $B_{11}$ &  $B_{21}$ & $B_{12}$ & $B_{22}$\\
 \hline
 Factors    $\downarrow$  &$A_0$& 00  11 & 00  22 &  00  11 & 00  22 & 00  11 & 00  22 \\
&$B_0$& 01  01 & 02  02 & 01  01 & 02  02 & 01  01 & 02  02  \\
             &$C_0$& 01  10 & 02  20 & 01  10 & 02  20 & 01  10 & 02  20 \\
             &$A_1$& 00  11 & 00 22 & 11  22 & 11  00 & 22  00 & 22  11 \\
             &$B_1$& 01  01 & 02 02 & 12 12  & 10  10 & 20  20 & 21 21 \\
             &$C_1$& 01  10 & 02 20 & 12 21 & 10 01 & 20  02 & 21  12 \\
      &$\tilde{A_1}$ & 00  11 & 22 00 & 11 22 & 00  11 & 22 00 & 11 22  \\
      &$\tilde{B_1}$ & 01  01 & 20 20 & 12 12 & 01 01 & 20  20 & 12 12 \\
      &$\tilde{C_1}$ & 01  10 & 20 02 & 12 21 & 01 10 & 20  02 & 12 21 \\
      \hline
            \end{tabular}
 \end{center}


\section{Inter-class orthogonal plans}

Inter-class orthogonal plans are defined in Bagchi (2019) in the
context of plans without any blocking factor. Here we extend the
definition  to the present context - the
orthogonality being through the block factor.

\begin{defi} \label{PIOTB} Let us consider a plan $\rho$. Suppose  the set of
 all factors of $\rho$ can be divided into several classes in such a way that
  if two factors belong to different classes, then they are
  orthogonal through the block factor. Such a plan  $\rho$ is
called a {\bf ``Plan Inter-class Orthogonal through the Blocks
(PIOTB)"} and the classes will be referred to as ``orthogonal
classes".\end{defi}

   We shall now proceed towards the construction of a series of PIOTBs. Using the
relation between orthogonal arrays of strength two and Hadamard
matrices, [see Theorem 7.5 in Hedayat, Sloane and Stuffken (1999),
for instance], we see that a $Q(n,n,2)$ exists whenever $n$ is the order of a Hadamard matrix.

\begin{theo}\label{PIOTBmn} Suppose  Hadamard matrices of orders $m$ and $n$ exist. Then, there exists a saturated PIOTB  ${\cal P}_{(m,n)}$ for a $2^{mn}$ experiment on $n$ blocks of size $m+1$ each. There are $n$ orthogonal classes of size $m$ each.
\end{theo}

{\bf Proof :} By hypothesis $Q_m = Q(m,m,2)$ exists. Let $R$ be the $m \times m+1$ array obtained by juxtaposing a column of all-ones to $Q_m$.
Let  $ {\cal P}_0$ be the plan for a $2^m$ experiment on a single block consisting of $m+1$ runs, which are the columns of $R$. Let us name the factors of $ {\cal P}_0$ as $A,B, \cdots M\}$. Note that the column added to $Q_m$ saves $A$ from being confounded with the block.

By hypothesis, $Q_n = Q(n,n,2)$ exists. Let ${\cal P}_{(m,n)} = Q_n \Diamond {\cal P}_0$. Clearly, ${\cal P}_{(m,n)}$ is an main effect plan for a $2^{mn}$ experiment with parameters as in the statement. By construction, no factor is confounded with the block factor.
Using Theorem \ref {PlanThruOA} and the property of $ {\cal P}_0$, we see that ${\cal P}_n$ is interclass orthogonal with orthogonal classes $\{A_i, B_i, \cdots M_i\},\; 1 \leq i \leq n$ (recall Notation \ref {allFactors}). Hence the result. $\Box$

We now present the plans ${\cal P}_{(4,4)}$.

\begin{center}
{\bf  Table 6.1 : The plan $ {\cal P}_{4,4}$}

\vskip5pt

\begin{tabular}{ll|c|c|c|c}
Blocks &$ \rightarrow$ &   $B_{1}$  &$B_1$ & $B_2$ &  $B_3$ \\
 \hline
 Factors  $\downarrow$
             &$A_1$& 00 00 1 & 00 00 1 &00 00 1 &00 00 1 \\
             &$B_1$& 00 11 1 & 00 11 1 &00 11 1 &00 11 1 \\
             &$C_1$& 01 01 1 & 01 01 1 &01 01 1 & 01 01 1\\
             &$D_1$& 01 10 1 & 01 10 1 &01 10 1 &01 10  1\\  \hline
             &$A_2$& 00 00 1 & 00 00 1 & 11 11 0 & 11 11 0 \\
             &$B_2$& 00 11 1 & 00 11 1 & 11 00 0 & 11 00 0 \\
             &$C_2$& 01 01 1 & 01 01 1 & 10 10 0 & 10 10 0\\
             &$D_2$& 01 10 1 & 01 10 1 & 10 01 0 &10 01 0\\\hline
             &$A_3$& 00 00 1 & 11 11 0 &00 00 1 &11 11 0 \\
             &$B_3$& 00 11 1 & 11 00 0 &00 11 1 &11 11 0 \\
             &$C_3$& 01 01 1 & 10 10 0 &01 01 1 &10 10 0\\
             &$D_3$& 01 10 1 & 10 01 0 &01 10 1 &10  010\\ \hline
             &$A_4$& 00 00 1 & 11 11 0 &11 11 0 & 00 00 1 \\
             &$B_4$& 00 11 1 & 11 00 0 &11 00 0 &  00 11 1 \\
             &$C_4$& 01 01 1 & 10 10 0 &10 10 0& 01 01 1 \\
             &$D_4$& 01 10 1 & 10 01 0&10  01 0&01  10 1
           \\\hline
            \end{tabular}
 \end{center}

There are four orthogonal classes, which are $\{A_i, B_i, C_i, D_i\}, i = 1, 2,3,4$.

\vspace{.5em}

Finally, we present a  PIOTB  for three-level factors.

\begin{theo}   A saturated PIOTB  exists for a $3^6$
experiment on four blocks of size four each.
\end{theo}
\vspace{.5em}

{\bf Proof :}  Consider the following plan ${\cal P}$ . It is easy
to see that it is a PIOTB with non-orthgonal classes
 $\{P_1,P_2\}, \: P = A,B,C$.

 \begin{center}
 {\bf  Table 6.2 :  Plan $ {\cal P}$}

\vskip5pt

\begin{tabular}{ll|c|c|c|c}
Blocks &$ \rightarrow$ &   $B_{1}$  &$B_2$ & $B_3$ &  $B_4$ \\
 \hline
 Factors  $\downarrow$   &$A_1$& 00 12 & 00 21 & 00 12 & 00 21 \\
 &$B_1$& 01 02 & 02 01 & 10 20 & 20 10 \\
             &$C_1$& 01 20 & 02 10 & 02 10 &  01 20 \\\hline
             &$A_{2}$& 01 01& 02 02& 01 01 & 02 02 \\
             &$B_2$& 01  10 & 02 20 &10  01& 20 02 \\
             &$C_2$& 00 11 & 00 22 & 11 00 & 22 00 \\
             \hline
                         \end{tabular}

                          \end{center}

{\bf Remark 6.1:} A POTB for a $4^4$ experiment on 4 blocks of size
4 is well-known [can be obtained by treating a row of
OA(16,5,4,2) as the block factor]. By collapsing two of the levels of each factor to one level
one gets a POTB for a $3^4$ experiment on the same set up.
 Allowing non-orthogonality we have been able to
accommodate two more three-level factors, making it saturated.

\section{References}
\begin{enumerate}
\item  Bagchi, S. (2010). Main effect plans orthogonal through the
block factor. Technometrics, vol. 52, p : 243-249.

\item  Bagchi, S. (2019). Inter-class orthogonal main effect plans for
asymmetrical experiments. Sankhya,  vol. 81-B,  p : 93-122.

\item  Bagchi, S. and Bagchi, B. (2020).  Aspects of optimality of plans orthogonal through other factors. Submitted.

\item Bose, M. and Bagchi, B. (2007). Optimal main effect plans in blocks of small size. Statist. Probab. Lett.,
vol. 77, p : 142-147.

\item Chen, X.P., JG Lin, J.G., Yang, J.F. and Wang, H.X (2015). Construction of main effects plans orthogonal through the block factor,
Statist. Probab. Lett. , vol. 106,  p :  58-64.

\item Das, A. and Dey, A. (2004). Optimal main effect plans with nonorthogonal
blocks. Sankhya, vol. 66, p : 378-384.

\item Hall, M. (1986). Combinatorial Theory. Wiley-interscince, New York.

\item Hedayat, A.S., Sloan, N.J.A. and Stufken, J. (1999). Orthogonal
arrays, Theory and Applications, Springer Series in Statistics.

\item Jacroux, Mike (2011). On the D-optimality of orthogonal and nonorthogonal blocked main effects plans. Statist. Probab. Lett. Vol. 81 ,  p:  116-120.

\item Jacroux, Mike (2011). On the D-optimality of nonorthogonal blocked main effects plans. Sankhya B, vol.  73,  p:   62-69.

\item Jacroux, Mike (2013). A note on the optimality of 2-level main effects plans in blocks of odd size. Statist. Probab. Lett. , vol. 83,  p:   1163-1166.

\item Jacroux, Mike, Kealy-Dichone, Bonni (2014). On the E-optimality of blocked main effects plans when $n  \equiv 3 \pmod 4$. Statist. Probab. Lett., vol.  87 ,  p:  143-148.

  \item Jacroux, Mike, Kealy-Dichone, Bonni (2015). On the E-optimality of blocked main effects plans when  $n \equiv 2 \pmod 4$. Sankhya B vol 77 ,  p:   165-174.
\item Jacroux, Mike, Jacroux, Tom  (2016).  On the E-optimality of blocked main effects plans when  $n \equiv 1 \pmod 4$. Comm. Statist. Theory Methods, vol. 45 ,  p:   5584-5589.

 \item Jacroux, Mike, Kealy-Dichone, Bonni  (2017). On the E-optimality of blocked main effects plans in blocks of different sizes. Comm. Statist. Theory Methods, vol.  46,  p:    2132-2138

\item Kiefer, J. (1975). Construction and optimality of generalized Youden designs. In: Srivastava,
J.N. (ed) A survey of statistical design and linear models. North-Holland, Amsterdam,
p : 333 - 353.

\item Morgan, J.P. and Uddin, N. (1996). Optimal blocked main effect
plans with nested rows and columns and related designs. Ann. Stat.
vol. 24, p : 1185-1208.

\item Mukerjee, R., Dey, A. and Chatterjee, K. (2002). Optimal
main effect plans with non-orthogonal blocking. Biometrika, 89, p : 225-229.

\item  Preece, D.A. (1966). Some balanced incomplete block designs
for two sets of treatment. Biometrika 53, p : 497-506.

\item Rees, D.H. and Preece, D.A. (1999). Perfect Graeco-Latin
balanced incomplete block designs. Disc. Math. vol.197/198, p : 691-712.

\item Rao, C.R. (1946). On Hypercubes of strength d and a system of confounding in factorial experiments Bull. Cal. Math. Soc., 38, p: 67.
1946

\item SahaRay, R., Dutta, G. (2016). On the Optimality of Blocked Main Effects Plans. International Scholarly and Scientific Research and Innovation, 10, p : 583-586.

\item Seberry, Jennifer, (1979). A note on orthogonal Graeco-Latin
designs. Ars. Combin. vol. 8, p : 85-94.

\item Shah, K.R. and Sinha, B.K. (1989). Theory of optimal
designs, Lecture notes in Stat., vol. 54, Springer-Verlag, Berlin.

\item Street, D.J. (1981). Graeco-Latin and nested row and column
designs. In Com. Math. VIII, Proc. 8th Austr. Conf. Comb. Math.,
Lecture notes in Math., vol. 884, Springer, Berlin. p : 304-313.

\item Takeuchi, K. (1961). On the optimality of certain type of PBIB
designs. Rep. Stat. Appl. Un. Jpn. Sci. Eng. vol. 8. p : 140-145.

\item Wang, P.C. (2004). Designing two-level fractional factorial
experiments in blocks of size two. Sankhya, vol. 66, p : 327.
\end{enumerate}

* Foot note : The author is retired from Indian Statistical Institute, Bangalore Center.
\end{document}